# Vortex pinning and dynamics in high performance $Sr_{0.6}K_{0.4}Fe_2As_2$ superconductor

Chiheng Dong,[1] He Lin,[1] He Huang,[1] Chao Yao,[1] Xianping Zhang,[1] Dongliang Wang,[1] Qianjun Zhang,[1] Yanwei Ma,[1,*] Satoshi Awaji[2] and Kazuo Watanabe[2]

1. Key Laboratory of Applied Superconductivity, Institute of Electrical Engineering, Chinese Academy of Sciences, Beijing 100190, People's Republic of China
2. High Field Laboratory for Superconducting Materials, Institute for Materials Research, Tohoku University, Sendai 980-8577, Japan

*email: ywma@mail.iee.ac.cn

**Abstract:**

We have studied vortex pinning and dynamics in a $Sr_{0.6}K_{0.4}Fe_2As_2$ superconducting tape with critical current density $J_c$~0.1 MA/cm$^2$ at 4.2 K and 10 T. It is found that grain boundary pinning is dominant in the vortex pinning mechanism. Furthermore, we observe large density of dislocations which can also serve as effective pinning centers. We find that the temperature dependence of critical current density is in agreement with the model of vortices pinned via spatial fluctuation of charge carrier mean free path. Magnetic relaxation measurement indicates that the magnetization depends on time in a logarithmic way. The relaxation rate in the low and intermediate temperature region is small, and it exhibits a weak temperature and field dependence. A crossover from elastic creep to plastic creep regime is observed. Finally, we conclude a vortex phase diagram for the high performance $Sr_{0.6}K_{0.4}Fe_2As_2$ superconducting tape.

# Introduction

The discovery of iron based superconductors[1] (IBSCs) encourages vast investigation into their physical properties and possible practical applications. The IBSCs are different from other applicable superconductors because of the high superconducting transition temperature $T_c$, high upper critical field[2] $H_{c2}$ and small anisotropy parameter $\gamma$. The critical current density $J_c$ in iron-based single crystals[3] and films[4] has already surpassed 1 MA/cm$^2$, indicating promising prospect of application in electronic devices. For strong current application, powder in tube (PIT) method maybe the most economical way to fabricate iron based superconducting wires and tapes. However, the incipient polycrystalline form of IBSCs suffers from two fatal defects: poor connectivity and weak link effect. The second phases, such as FeAs, amorphous and oxide layers[5-7], existing between grains significantly degrade the intergrain current. Pores and cracks decrease the effective current-carrying cross-sectional area and hence suppress $J_c$. Further studies on the epitaxial films indicated that the intergrain critical current density $J_c^{gb}$ decreases rapidly[8,9] when the misoriented angle of grain boundary is larger than 9°. Moreover, the $J_c^{gb}$ with large misoriented angle declines sharply even in a small magnetic field.

To overcome these obstacles, varieties of strategies, *e.g.* metal addiction[10-12], rolling texture[13,14], hot isostatic pressing[15], uniaxial pressing[16-18], were applied to increase the connectivity, density and texture of the superconducting core. Among them, the hot pressing (HP) method is considered as a very efficient way to improve $J_c$. The critical current density at 4.2 K and 10 T of the hot pressed tapes has surpassed 0.1 MA/cm$^2$. It is found that the number of voids and residual cracks greatly decreases after hot pressing, and the grain boundaries are clear without any second phases. Electron backscatter diffraction measurements[19] indicate that the dominant orientation of the grains is (00*l*), and many misorientation angles of the grain boundaries are smaller than 5°. All these results suggest that the superconducting core of the hot-pressed iron based superconducting tape is highly textured with good connectivity.

Besides the connectivity and texture, vortex pinning and motion is another important factor that influences $J_c$. Studies on the iron based single crystals indicated that the small sized normal cores[20] or fluctuating magnetic/structural domains[21] are responsible for vortex pining. Relaxation of magnetization suggested a moderate vortex creep rates and relatively small characteristic pinning energy[22]. However, all these results are based on the single crystalline sample and inapplicable to the polycrystalline superconducting tapes. In this paper, we will discuss vortex pinning mechanism and dynamics for the high performance $Sr_{0.6}K_{0.4}Fe_2As_2$ tape fabricated by the hot pressing method. It is found that the grain boundaries are the predominant pinning centers. Furthermore, we observe large density of dislocations via transmission electron microscopy, which also participate in vortex pinning. Moreover, the accordance of the experimental data and the $\delta l$ pinning curve indicates that the fluctuation of mean free path is the main pinning source. We additionally demonstrate a small vortex relaxation rate with weak temperature and field dependence. A crossover from elastic vortex creep to plastic creep regime is observed. Finally, we conclude a vortex phase diagram for the high performance $Sr_{0.6}K_{0.4}Fe_2As_2$ tapes.

**Experiment**

The Sr-122 precursor was prepared by the solid state reaction method. Sr fillings (99.9 %), K pieces (99.95 %), As (99.95 %) and Fe (99.99 %) powders were loaded to the ball milling jar according to the nominal composition $Sr_{0.6}K_{0.5}Fe_2As_2$. We added excess K to compensate the loss during heat treatment. After 10 hours of ball milling, the thoroughly mixed powders were loaded into a Nb tube which is then inserted into an iron tube, welded shut and finally heated to 900 ℃ for 35 hours. The as-sintered precursor was ground and further packed into an Ag tube with the outer diameter (OD) of 8 mm and inter diameter (ID) of 5 mm. The Ag tube were swaged and drawn into an Ag wire with the OD of 1.9 mm and then flat rolled into a tape with 0.4 mm thickness. Finally, the tapes were hot-pressed under ~30 Mpa at 880 ℃. We measured the transport critical current $I_c$ at 4.2 K under a magnetic field range from 0 to 14 T at the High Field Laboratory for Superconducting Materials (HFLSM) at Sendai. We applied a four-probe method and determined $I_c$ with a criterion of 1 $\mu$V/cm. To further study the microstructure and magnetic properties, we peeled off the sheath and performed measurements on the superconducting core. The composition of the superconducting core determined by Energy Dispersive X-ray Spectroscopy is close to the nominal composition $Sr_{0.6}K_{0.4}Fe_2Se_2$. We therefore use this formula throughout the paper. X-ray diffraction confirms the high purity of our sample. The microstructure of the plane of the superconducting core was examined on a JEOL JEM-2100F transmission electron microscope (TEM). Magnetization of the superconducting core with dimensions of $2\times3\times0.1$ mm$^3$ was measured by a SQUID-VSM on a Magnetic Property Measurement System (MPMS3, Quantum Design). Vortex-dynamics was investigated by measuring the isothermic magnetization, M *vs.* time curves, over a period of 1 hour. Resistivity up to 14 T was measured on the same sample via a four-probe method on a Physical Property Measurement System (PPMS).

**Results and discussion**

Fig.1 shows the field dependence of the transport critical current density $J_c^{trans}$ at 4.2 K. $J_c^{trans}$ at 2 T achieves 0.15 MA/cm$^2$ ($I_c$~450 A) and gradually decreases with magnetic field. At 10 T, $J_c^{trans}$ approaches 0.1 MA/cm$^2$ ($I_c$~300 A), manifesting a robust field dependence. The inset (a) shows the temperature dependence of susceptibility with field perpendicular to the tape plane. A diamagnetic signal begins at $T_c$~35 K and saturates near 25 K. The magnetization with a field cooling process is independent on temperature, indicating strong vortex pinning. Inset (b) shows the temperature dependence of resistivity. At normal state, the resistivity exhibits a metallic temperature dependence, following a sharp superconducting transition at 36 K. The transition width is $\Delta T_c$~0.7 K, manifesting high homogeneity of the sample. The resistivity at 37 K is 0.088 mΩ•cm. The residual resistivity ratio is RRR=$\rho$(250 K)/ $\rho$(37 K)=4.81. For comparison, the $\rho$(37 K) and RRR value of the $Sr_{0.6}K_{0.4}Fe_2As_2$ single crystal[23] are 0.073 mΩ•cm and 5.28, respectively. The values of resistivity of our sample are close to that of the single crystal, suggesting that the hot pressed sample is of good connectivity.

Fig.2 (a) shows the isothermal hysteresis loops between 5 K and 30 K. The magnetic field is perpendicular to the tape plane. The symmetric shape of the hysteresis loop manifests that the vortex pinning is mostly due to bulk pinning rather than surface barrier. Below 20 K, the highest magnetic field attainable in our instrument (7 T) is not sufficient to achieve the closure of the hysteresis loop, indicating a high irreversibility field. Distinct from FeAs-122 single crystals, there is no trace of second magnetization peak (fishtail effect) in the whole temperature range.

We calculate the in-plane magnetic critical current density $J_c^{mag}$ for a sample with rectangular shape with dimension $c<a<b$ via the Bean model:

$$J_c^{mag} = \frac{20\Delta M}{a(1-a/3b)}, \quad (1)$$

where $J_c^{mag}$ is the in-plane critical current density in A/cm$^2$, $a$ and $b$ ($a<b$) are the width and length of the sample in cm, $\Delta M$ is the difference between the magnetization values for increasing and decreasing field in emu/cm$^3$. The field dependence of the calculated $J_c^{mag}$ is shown in Fig.2 (b). At 0 T and 5 K, the critical current density approaches 0.2 MA/cm$^2$. With field increases, there is a remarkable retention of $J_c$ at high fields. The $J_c^{mag}(B)$ curve nearly overlaps with the $J_c^{trans}(B)$ curve, indicating that the $J_c^{mag}$ is mainly contributed by the global current and the granularity of this sample can be neglected. However, $J_c^{mag}$ decreases more quickly when temperature increases above 20 K, exhibiting a convex curvature which is quite common in high temperature superconductors.

We plot $J_c^{1/2}(B)^{1/4}$ as a function of $B$ to evaluate the irreversibility field H$_{irr}$, as shown in the inset of Fig.2 (c). It is found that the Kramer plots present a wide linear behavior. The H$_{irr}$ is estimated as the extrapolated zero $J_c^{1/2}(B)^{1/4}$ value[24]. In order to study the vortex pinning mechanism in the high performance Sr$_{0.6}$K$_{0.4}$Fe$_2$As$_2$ tapes, we plot the normalized vortex pinning force $f=F_p/F_p^{max}$ as a function of the reduced field $h=H/H_{irr}$, as shown in Fig. 2 (c). According to the Dew-Hughes model[25], the normalized vortex pinning forces $f=F_p/F_p^{max}$ at different temperatures will collapse into one unified curve described by $f=Ah^p(1-h)^q$ if there is a dominant pinning mechanism. As shown in Fig.2 (c), the normalized curves of $f(h, T)$ at different temperatures overlap and present the same scaling law, demonstrating that the pinning mechanism is same within the temperature range. We fit the $f(h, T)$ curves with the function $f=Ah^p(1-h)^q$ and estimate $p$=0.64, $q$=2.24. The calculated peak position is $h_{max}$=$p/(p+q)$=0.22, agrees well with the observed $h_{max}$ value.

For type II superconductors, vortices interact with pinning centers via the spatial variations of $T_c$ ($\delta T_c$ pinning) or by the scattering of charge carriers with mean free path $l$ near defects ($\delta l$ pinning). For the $\delta l$ pinning, $h_{max}$~0.33 and 0.2 correspond to the point pinning and surface pinning, respectively. While for the $\delta T_c$ pinning, $h_{max}$ locates at 0.67, 0.6 and 0.5 for the point, surface and body pinning, respectively. In our case, $h_{max}$ is close to 0.2 and we propose that grain boundary pinning is dominant due to the polycrystalline nature, which is similar with that of the $Nb_3Sn$[26], $PbMo_6S_8$[27] and $MgB_2$[28] superconductors. The inset in (d) is a TEM image of typical dislocations on the tape plane, which can be seen in many grains of the sample. It is known that density of dislocations in material increases with deformation stress in the form: $\tau \propto \sqrt{\rho}$, where $\rho$ is the density of dislocations, $\tau$ is the deformation stress. We suggest that hot pressing process greatly enhances the deformation stress and consequently increases the density of dislocations. These dislocations can interact with one vortex line when the dislocation is lying parallel to the local field, or with several vortex lines when there is an angle between the dislocation and the local field. Thus, we believe that the widely existed dislocations also contribute to the vortex pinning.

In order to elucidate the origin of pinning, we analyze the temperature dependence of the normalized critical current density $J_c$. According to the theory proposed by Griessen[29] *et al.*, in the case of $\delta l$-type weak pinning, the normalized $J_c$ can be described by the following expression:

$$\frac{J_c(t)}{J_c(0)} \propto (1-t^2)^{5/2}(1+t^2)^{-1/2}. \qquad (2)$$

For the $\delta T_c$ pinning, it is described as:

$$\frac{J_c(t)}{J_c(0)} \propto (1-t^2)^{7/6}(1+t^2)^{5/6} \qquad (3)$$

where $t=T/T_c$. As shown in Fig.2 (d), the experimental results at 0, 0.5 and 1 T are very close to the theoretical curve of $\delta l$ pinning. It indicates that the vortex pinning in the hot-pressed $Sr_{0.6}K_{0.4}Fe_2As_2$ tape originates from the spatial variation of mean free path, which is similar to that of the silicone-oil-doped $MgB_2$ bulk samples[30].

For type-II superconductors, attractive interactions between vortices and pinning centers prevent movement of vortices. However, thermal activation, quantum tunneling[31] or other external activation[32] can make vortices or vortex bundles escape from pinning wells. The nonequilibrium configuration of vortices will relax and lead to a redistribution of current loops and hence a change of magnetization *vs.* time. Therefore, measurement of relaxation of irreversible magnetization is important to study vortex pinning and creep, vortex phase diagram as well as fish tail effect.

Fig. 3 shows the field and temperature dependence of the normalized magnetization relaxation rate $S = d\ln(-M)/d\ln(t)$. The inset in (c) depicts the time dependence of the magnetization at 1 T on the log-log plot. We find a very slow relaxation, *e.g.*, there is only 1.3 % change of magnetization at 10 K and 1 T within 4000 s. In addition, we observe a linear dependence of ln(-M) on ln(t), which is consistent with the model of thermally activated vortex motion. Assuming thermal activation over the effective pinning barrier U(*j*, T, B), the electric field induced by the vortex motion is

$$E = v_0 B e^{-\frac{U(j,T,B)}{k_B T}}, \qquad (4)$$

Where $k_B$ is the Boltzmann constant, $v_0$ is the attempt hopping velocity. The effective pinning barrier can be described by the "interpolation formula" which covers all known functional forms of U(*j*, T, B)[33]:

$$U(j, T, B) = \frac{U_c(T, B)}{\mu(T, B)} \left[ \left( \frac{j_c(T, B)}{j(T, B)} \right)^{\mu(T, B)} - 1 \right] \qquad (5)$$

, where $\mu$ is the glassy exponent depending on the dimensionality and vortex creep regimes, $U_c$ is the characteristic pinning energy, $J_c$ is the unrelaxed critical current density. According to Schnack *et al.*[34] and Jirsa *et al.*[35], the dynamical relaxation rate (Q=dln*j*/dln*E*) should be same as the normalized relaxation rate (S=dlnM/dln*t*) measured by the conventional method. With this definition, we can derive the following equation from Eqs. (4) and (5)

$$\frac{T}{S(T,B)} = \frac{U_c(T,B)}{k_B} + \mu(T,B)CT \tag{6}$$

, where $C=\ln[2v_0B/(ldB/dt)]$ is a parameter with weak temperature dependence, $l$ is the sample's lateral dimension.

Fig. 3. (a) shows the field dependence of the normalized relaxation rates below 10 K. The relaxation rates are below 0.01 and exhibit a nonmonotonic field dependence. Below 0.5 T, the relaxation rates increase quickly with field, and then reach a peak at 0.5-1 T. After the peak, the relaxation rates decrease with magnetic field, go to a trough at ~2-3 T, and finally increase monotonically with magnetic field. This field dependent behavior of S is very similar to that of the Ba(Fe$_{0.92}$Co$_{0.08}$)$_2$As$_2$ single crystals[22], which is explained by the crossover between different regimes of vortex dynamics. It is noteworthy that the relaxation rate below 10 K is one order smaller than that of the Ba(Fe$_{1-x}$Co$_x$)$_2$As$_2$[22,36,37] single crystals and cuprates[33,38], but comparable to that of the LiFeAs[39] and irradiated Ba(Fe$_{0.93}$Co$_{0.07}$)$_2$As$_2$ single crystals[40]. The rather small values of relaxation rates suggest strong vortex pinning at low temperature. As shown in Fig.3 (b), the S values at 10 K and 15 K exhibit a weak field dependence below 3 T. While above 25 K, the relaxation rate increases quickly with field.

Fig.3 (c) shows the temperature dependence of the normalized relaxation rate. The S-T curves exhibit a plateau in the low and intermediate temperature region, the values of which fall into a narrow range from 0.001 to 0.005. With temperature further increases, the relaxation rate starts to diverge. This sudden increase of relaxation rate may correspond to a crossover between different vortex creep regimes. Moreover, the temperature at which S diverges decreases with increasing magnetic field. The plateau like behavior of S(T) curve is very similar to that of the SmFeAsO$_{0.9}$F$_{0.1}$[41], BaFe$_{1.92}$Co$_{0.08}$As$_2$[22] and some copper-based superconductors[38,42]. It can be well explained by the collective pinning model rather than the single vortex creep with rigid length as predicted by the Kim-Anderson model.

The temperature dependence of T/S at different fields are shown in Fig. 3 (d). At low temperature, T/S values increase linearly with T. According to Eqs. (6), the intercepts of T/S *vs.* T curves with the ordinate provide the characteristic pinning energy at 0 K. We extrapolate the T/S *vs.* T curve at 1 T linearly to 0 K and estimate $U_c(0)$ to be 174 K. This value is close to that of the $Tl_2Ba_2CaCu_2O_8$ thin films ($U_c(0)\sim 150$ K)[38] but larger than that of the optimal doped $Ba(Fe_{1-x}Co_x)_2As_2$ single crystal ($U_c(0)\sim 98$ K at 0.5 T)[22] and $SmFeAsO_{0.9}F_{0.1}$ polycrystals ($U_c(0)\sim 40$ K at 1 T)[41]. The large characteristic pinning energy indicates strong vortex pinning in the hot-pressed $Sr_{0.6}K_{0.4}Fe_2As_2$ tape. After the initial ascent, T/S curves reach a peak and decrease afterwards, universally exhibiting a peak shape behavior. We define a crossover temperature $T_{cr}$ at which $d(T/S)/dT=0$. For B=0.5 T, the T/S *vs.* T curve gives a positive slope at low temperature and changes to negative when temperature is above $T_{cr}\sim 16.5$ K. The crossover temperature shows strongly field dependence, and moves down to the lower temperature with increasing field.

According to Ref.[43], the $\mu$ value can be obtained from Eqs. (5) and (6)

$$\mu = -S \frac{d^2 \ln E}{d \ln j^2}. \qquad (7)$$

From this equation, we can derive the curvature of the lnE *vs.* lnJ curve from the sign of $\mu$. A positive $\mu$ corresponds to a negative curvature of the lnE-lnJ curve which is expected by the elastic creep of vortex. While a negative $\mu$ corresponds to positive curvature and plastic motion of vortex. As shown in Fig. 3 (d), T/S *vs.* T curves show positive slope below $T_{cr}$, corresponding to positive $\mu$ and elastic vortex motion. When temperature increases above $T_{cr}$, the curves suggest a negative $\mu$ and vortex motion transits to the plastic creep regime. The peak shaped T/S *vs.* T curve and the transition between different vortex motion regimes are also observed in the optimal doped $Ba(Fe_{1-x}Co_x)_2As_2$ single crystals[22], $Tl_2Ba_2CaCu_2O_8$ and $YBa_2Cu_3O_{7-\delta}$ thin films[38,42].

On the basis of the obtained data, we construct the vortex phase diagram for the high performance $Sr_{0.6}K_{0.4}Fe_2As_2$ tape in Fig. 4. The upper critical field $H_{c2}$ and irreversibility field $H_{irr}$ are derived from the magneto-resistivity and Kramer plot, respectively. $H_{cross}$ is defined as the crossover field from elastic to the plastic creep regime. All the characteristic fields exhibit strong temperature dependence and divide the mixed state into three regions. At low temperature and field, the vortex moves in an elastic way with a small relaxation rate. With increasing temperature and field, the vortex system enters into a plastic creep region in which the relaxation rate undergoes a fast increase. Above the $H_{irr}(T)$ line, the vortex is highly mobile and the superconducting current disappears. This phase diagram is analogous to that of other high temperature superconductors[22,36,38,39,44] except that $T_{cr}$ at zero field is far below $T_c$ ($T_{cr}(0\,T)/T_c \sim 0.6$). This behavior is also observed in the underdoped $Ca_{1-x}Na_xFe_2As_2$ single crystal[45], whose $T_{cr}(0\,T)/T_c$ ratio is coincidently the same as ours.

In the elastic creep area as shown in Fig. 4, the magnetization relaxation rate remains a small value ($\sim 3 \times 10^{-3}$). The characteristic pinning energy $U_c$ depends on the magnetic field B in the form: $U_c^{ela} \propto B^v$, where $v$ is positive and depends on the specific pinning regime. When the vortex system enters into the plastic regime which occupies a large area of the phase diagram, the magnetization relaxation rate quickly increases to above 0.02. The characteristic pinning energy in this area can be written as $U_c^{pla} \propto (\phi_0^2 d)/(4\pi\mu_0 \lambda^2 \gamma)$, where $\phi_0$, $\lambda$, $\gamma$ and $d \propto \sqrt{\phi_0/B}$ is the vortex quantum, penetration depth, anisotropy parameter and inter-vortex distance, respectively. So the pinning energy decreases with field in the form $U_c^{pla} \propto B^{-0.5}$. Therefore, we can expect higher and more robust field dependence of $J_c$ in the elastic creep regime. Within the liquid helium temperature region, which is in the elastic creep area, $J_c$ exhibits weak field dependence as shown in Fig. 1 and Fig. 2 (b). While in the liquid hydrogen temperature region, which is in the plastic creep area, $J_c$ quickly decreases from $10^5$ A/cm$^2$ with increasing magnetic field as shown in Fig. 2 (b). As a result, if we want to apply the $Sr_{0.6}K_{0.4}Fe_2As_2$ superconducting tapes in the low field magnets (<10 T) operating at intermediate temperature (*e.g.* cryocooler or liquid hydrogen), new technologies are required to expand the elastic creep area in the vortex phase diagram.

According to our results, the grain boundaries act as effective pinning sites in the hot-pressed $Sr_{0.6}K_{0.4}Fe_2As_2$ tape. This is quite similar to the Nb-Ti, $Nb_3Sn$ and $MgB_2$ superconducting wires, in which grain boundaries are beneficial vortex pinning sites without blocking intergranular superconducting current[28]. Pande *et al.* proposed a model of vortex pinning by grain boundaries in type-II superconductors and confirmed the experimental $F_p \propto 1/D$ relationship, where D is the particle size[46]. Martínez *et al.* studied the relationship between the vortex pinning force and the grain size in $MgB_2$. He found that $F_p$ was low and independent on the grain size for the samples with bad grain connectivity. While for the sample with good grain connectivity, the pinning force increases with decreasing grain size[47]. In our case, better recrystallization under high pressure make the number of voids, cracks and the proportion of large misorientation angle greatly decrease. Consequently, the connectivity of the hot-pressed $Sr_{0.6}K_{0.4}Fe_2As_2$ tapes greatly improves. We can therefore expect larger vortex pinning force for the samples with smaller grain size. Further experimental evidence, however, is still needed to elucidate the intrinsic relationship between $F_p$ and the grain size. On the other hand, it is suggested that point pinning is more efficient than surface pinning due to its smaller size comparable to the coherence length. As a result, artificially introducing point defects by means of irradiation and nanoparticle inclusion may be a feasible way to enhance the current carrying ability.

**Conclusions**

We have studied the vortex pinning mechanism and vortex dynamics in the high performance $Sr_{0.6}K_{0.4}Fe_2As_2$ tape which achieves $J_c \sim 10^5$ $A/cm^2$ at 4.2 K and 10 T. It is found that grain boundaries are the main pinning centers. In the meantime, the widely existed dislocations observed by TEM also contribute to the vortex pinning. The good agreement between the $J_c(T)$ curve and the $\delta l$ pinning theoretical curve indicates that the pinning originates from spatial variation of mean free path. Through the magnetic relaxation measurement, we discover a logarithmic dependence of magnetization on time. The small relaxation rate has a weak temperature and field dependence in the low and intermediate temperature region. Based on the results, we concluded a vortex phase diagram for the high performance $Sr_{0.6}K_{0.4}Fe_2As_2$ superconducting tapes. It is found that there is a crossover from elastic vortex creep to plastic vortex creep regime. We propose that decreasing grain size and introducing point defects might be beneficial to $J_c$ enhancement.

**Acknowledgements**

The authors are grateful to Kai Wang and Chen Li for their useful comments. This work is partially supported by the National Natural Science Foundation of China (Grant Nos. 51320105015 and 51402292), the Beijing Municipal Science and Technology Commission (Grant No. Z141100004214002), the Beijing Training Project for the Leading Talents in S & T (Grant No. Z151100000315001).

**Captions**

**Figure 1**
Field dependence of the transport critical current density $J_c$ for the hot-pressed $Sr_{0.6}K_{0.4}Fe_2As_2$ tape at 4.2 K. Inset (a) exhibits the temperature dependence of the susceptibility for the superconducting core with zero field cooling (ZFC) and field cooling (FC) procedures (H=30 Oe ⊥ tape plane). Inset (b) is the resistivity as a function of temperature.

**Figure 2**
(a) Isothermic M(H) curves at 5 K-30 K. The magnetic field is perpendicular to the tape plane. (b) Field dependence of the magnetic critical current density at 5 K-30 K. (c) Normalized vortex pinning force $f_p=F_p/F_p^{max}$ as a function of the reduced field $b=B/B_{irr}$. The dashed-dotted line is the fitting curve using $f_p=Ah^p(1-h)^q$. The inset shows the Kramer plot for T=20 K, 22.5 K, 25 K and 27.5 K. The dashed lines are linear fit to the curves. (d) Temperature dependence of the normalized critical current density at 0 T, 0.5 T and 1 T. The black and blue dashed-dotted lines represent the $\delta T_c$ and $\delta l$ pinning, respectively. The inset shows a TEM image of typical dislocations on the tape plane, which can be seen in many grains of the sample.

**Figure 3**
(a) and (b): Field dependence of the normalized magnetization relaxation rate at different temperatures. (c) Temperature dependence of the relaxation rate at various fields. Inset shows the time dependence of the magnetization at 1 T from 4 K to 25 K on the double logarithmic scales. (d) Temperature dependence of T/S at different fields.

**Figure 4**
The vortex phase diagram for the high performance $Sr_{0.6}K_{0.4}Fe_2As_2$ tape. The upper critical field $H_{c2}$ is determined by the criterion of the 90 % normal state resistivity; the irreversibility field is determined by the Kramer plot; the $H_{cross}(T)$ line is derived from the temperature dependence of T/S.

**Figure 1**

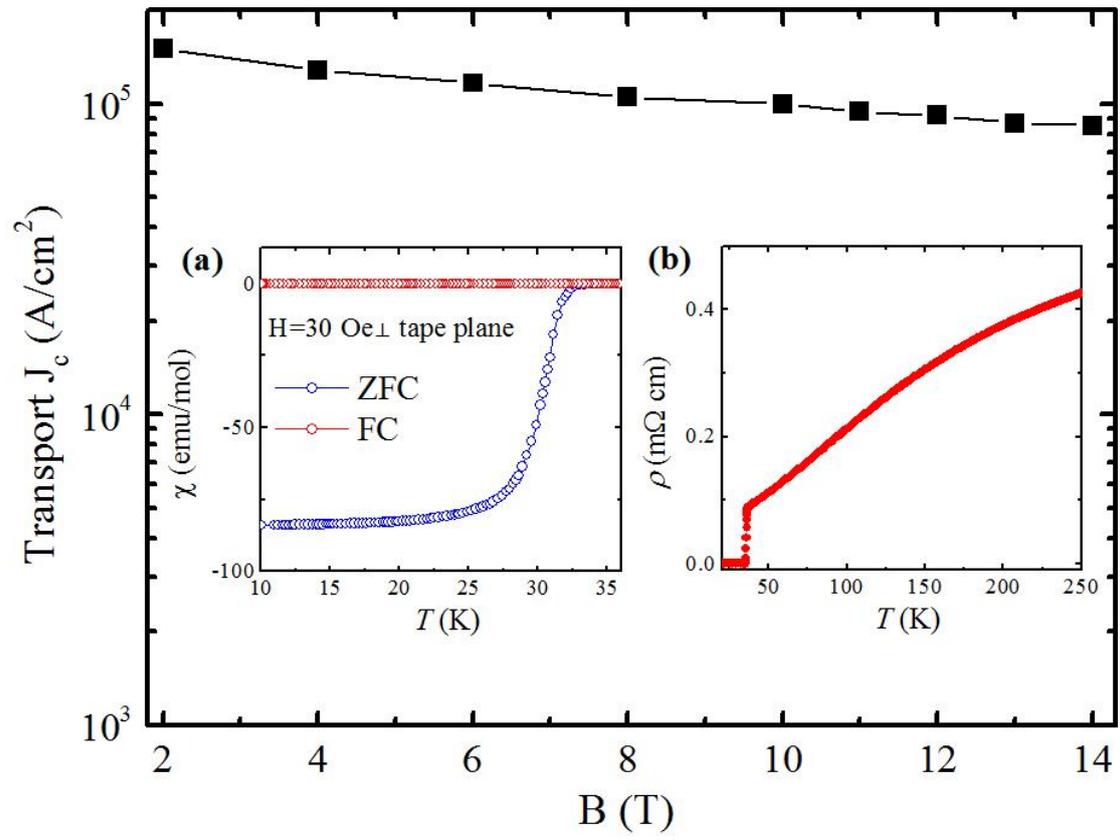

**Figure 2**

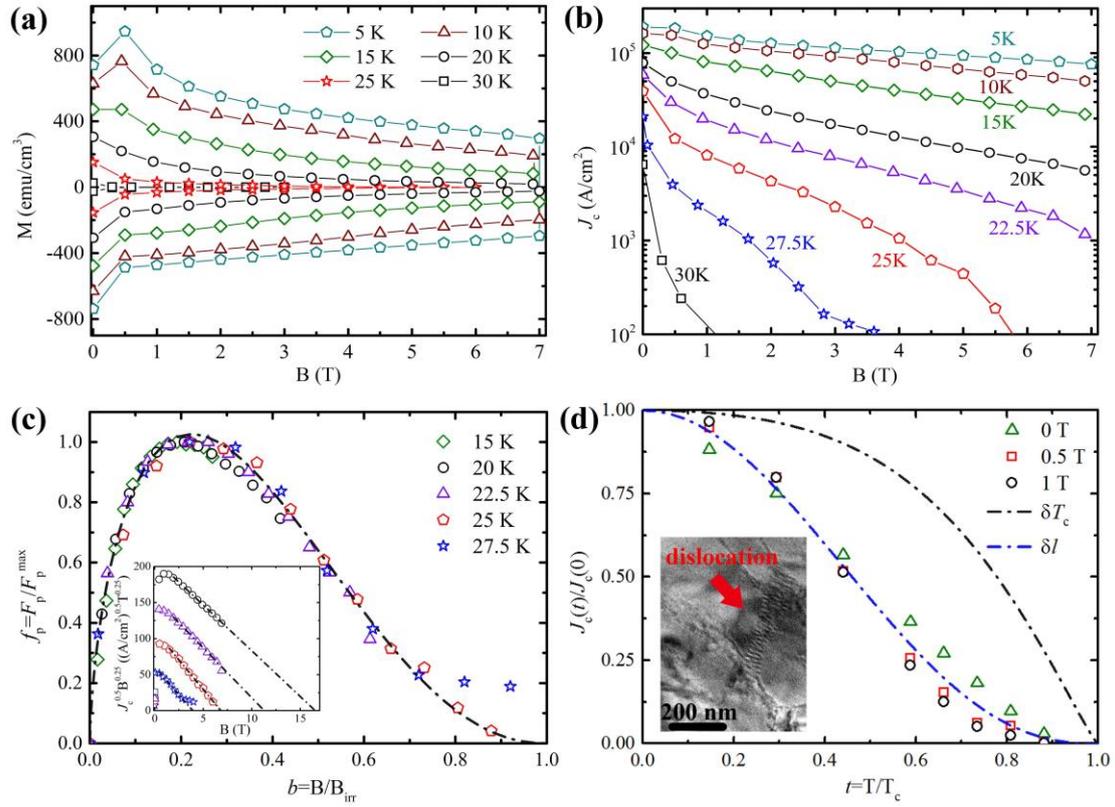

**Figure 3**

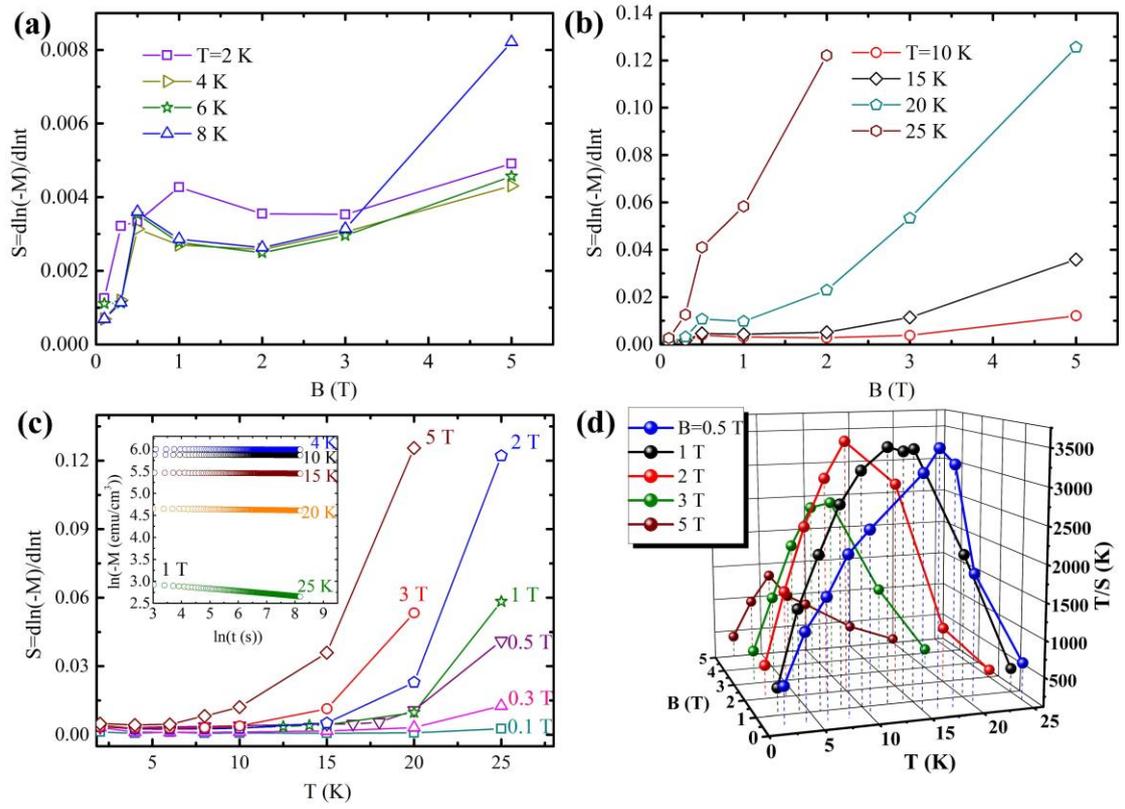

**Figure 4**

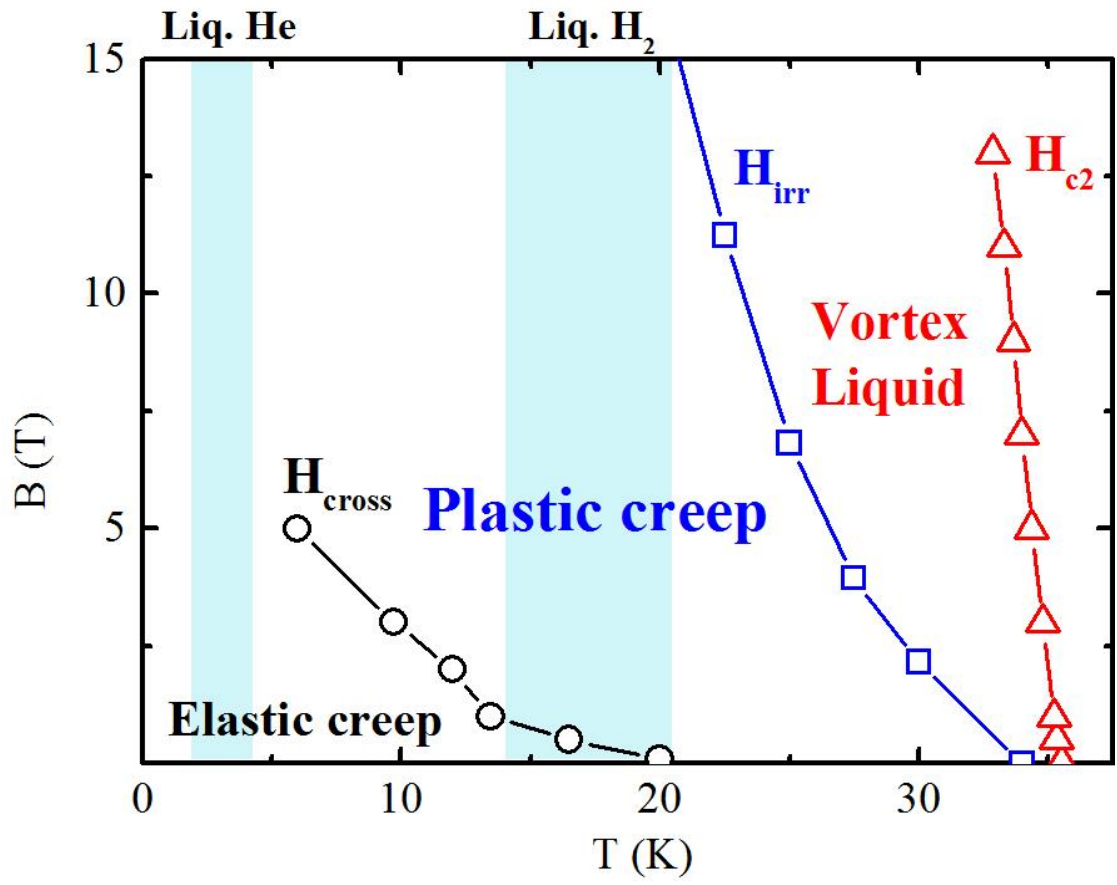